\documentclass[aps,prd,a4paper,superscriptaddress,nofootinbib,
showpacs,twocolumn,showkeys,amsfonts,amssymb,amsmath]{revtex4}

\usepackage{amssymb,latexsym}
\usepackage{amsmath, amsthm}
\usepackage{amscd}
\usepackage{times}
\usepackage{epsfig}
\usepackage{psfrag}
\usepackage{graphicx}
\usepackage{overpic}
\usepackage{booktabs,tabularx}

\newcommand{\be}{\begin{equation}}
\newcommand{\ee}{\end{equation}}
\newcommand{\bea}{\begin{eqnarray}}
\newcommand{\eea}{\end{eqnarray}}

\begin{document}
\title{Gravitational collapse of Hagedorn fluids}

\author{Daniele Malafarina} \email{daniele.malafarina@nu.edu.kz}
\affiliation{Department of Physics,
Nazarbayev University, 53 Kabanbay Batyr, 010000 Astana, Kazakhstan}

\swapnumbers

\begin{abstract}
We consider a toy model for relativistic collapse of an homogeneous perfect fluid that takes into account
an equation of state for high density matter, in the form of an Hagedorn phase, and semiclassical corrections in the strong field.
We show that collapse reaches a critical minimum size and then bounces.
We discuss the conditions for collapse to halt and form a compact object.
We argue that implications of models such as the one presented here are of great importance for astrophysics as they show
that black holes may not be the only final outcome of collapse of very massive stars.
\end{abstract}


\maketitle

\section{Introduction}

Relativistic gravitational collapse is at the foundation of black hole physics.
It is widely believed that the collapse of a sufficiently massive star will lead
inevitably to the formation of a black hole.
This idea is rooted in the simplest collapse model, developed by Oppenheimer and Snyder and independently by Datt (OSD) in 1939
\cite{OSD}.
The OSD model describes a spherical, non rotating, homogeneous matter cloud (made of pressureless, `dust', particles) that collapses under its own weight.
In the OSD model, after the boundary of the cloud passes the Schwarzschild radius, the black hole forms. All the matter eventually falls into the central singularity that remains hidden from far away observers.
Many analytical studies of relativistic spherical collapse were developed starting from the pioneering work of Oppenheimer and Snyder and Datt.
These classical models are solutions of Einstein's equations for physical matter sources such as dust, perfect fluids, fluids with only tangential pressures, with and without inhomogeneities
(see for example \cite{class}).
In more recent times some attention has been devoted to the study of collapse models that take into account corrections to general relativity at high densities.
These semiclassical models are solutions of Einstein's equations for effective matter sources that are composed of a physical part together with an unphysical part that describes modifications to general relativity in the strong field
(see for example \cite{semiclass}).

Classical models, satisfying standard energy conditions, generically develop singularities
\cite{sing}.
However, it is usually believed that singularities should not form in the real universe and that their appearance in solutions of Einstein's equations merely signals a breakdown of the theory in the strong field regime.
Therefore, in order to avoid the formation of singularities, at some stage during collapse either the matter model must
violate energy conditions or general relativity must not hold.
It was Wheeler who first recognized the importance of classical singularities as possible windows on a regime where quantum-gravity dominates.
Singularities can be avoided in semiclassical models where repulsive forces, arising at high densities, balance the gravitational attraction, and bouncing scenarios are favored.
For example in the last few years there have been several studies of collapse in semiclassical models inspired by bounces in Loop Quantum Cosmology (LQC) (see
\cite{bojowald}).
These bouncing models exhibit a minimum scale (related to the maximum allowed density, that is in turn related to the energy scale of quantum-gravity) that suggests the possibility of the existence of small compact remnants.
Exotic compact remnants as leftovers from gravitational collapse have been discussed for decades,
and while their theoretical properties have been widely studied their existence is still purely hypothetical.
There exist many proposed objects from gravastars
\cite{grava},
to quark stars
\cite{quark},
to boson stars
\cite{boson}
and Planck stars
\cite{planck}.

Here we present a simple toy model for collapse of an homogeneous perfect fluid that considers both the equation of state for high density as well as semiclassical corrections that occur in the strong field.
With respect to previously studied models the scenario presented here depends upon two parameters, which are in turn related to two different energy scales. One is the characteristic scale of the Hagedorn phase and the other is the energy scale of semi-classical corrections. Note that, depending on the chosen approach to modify General Relativity, the latter may or may not be related to the Planck scale. In the present work we use a standard semi-classical approach that comes from Loop Quantum Gravity (LQG).
The model is particularly appealing because all physical quantities are well defined, physically meaningful and well behaved.
The main result is that collapse reaches a critical size characterized by a maximum critical density and then bounces.
Therefore the collapsing phase that leads to the formation of a black hole is followed by an expanding phase that can be described
as a white hole solution. The main difference with the standard OSD collapse model comes from the behaviour of trapped surfaces.
Trapped surfaces initially develop similarly to the classical case but `evaporate' before collapse reaches the critical stage.
One main consequence of the existence of a critical scale is that there is a minimum radius below which the horizon does not form at
all. Such radius is related to the critical density and for quantum-gravitational effects is of the order of the Planck length.
Therefore, this model suggests the possibility of the existence of exotic compact remnants as leftovers from collapse.
The implications of such models for astrophysics are immediately clear:
A sufficiently massive star that collapses under its own gravity may not necessarily end up in a black hole.
In fact we show here that its core may produce an extremely dense, not very massive, exotic compact object while
the outer layers and most of its mass are ejected in an explosion that is powered at the level of the quantum-gravity scale.

The paper is organized as follows: In section \ref{pf} classical models for collapse of homogeneous perfect fluids are briefly reviewed. In section \ref{hagedorn} the Hagedorn equation of state (e.o.s.) is introduced and classical perfect fluid collapse with such e.o.s. is investigated.
In section \ref{semi} semiclassical collapse of the Hagedorn fluid is described.
Finally section \ref{disc} is devoted to discussing possible astrophysical implications of the model.
In the following we use geometrical units for which $G=c=1$.

\section{Perfect fluid collapse}\label{pf}
The OSD model can readily be extended to the case of collapse of an homogeneous perfect fluid sphere, with linear equation of state relating the energy density $\rho(t)$ to the pressure $p(t)$, in the form $p=k\rho$.
We use co-moving coordinates $\{r,t\}$, that can be thought of as coordinates attached to the infalling particles of the cloud, for which
the energy momentum tensor takes diagonal form as $T^{\mu\nu}=\text{diag}\{\rho,p,p,p\}$. Standard energy conditions must be satisfied by energy density and pressure. Typically we require the weak energy conditions (w.e.c.), that can be written as $\rho>0$ and $\rho+p\geq 0$ and imply $k\geq-1$. The speed of sound in the cloud $c_s$ is defined as $c_s^2=dp/d\rho=k$, so that for $c_s$ not to exceed the speed of light we must require $k\leq 1$. Finally in order to have positive pressures one should impose $k>0$. Note however that in certain cases negative values of $k$ may be considered. For example, this is the case of the dark energy e.o.s. which requires $k=-1$ and for which the density reduces to the cosmological constant.
The amount of matter enclosed within the radius $r$ at the time $t$ is described via the Misner-Sharp mass of the system $F(r,t)$, that for a homogeneous perfect fluid, in order to satisfy regularity requirements at the center, can be written as $F(r,t)=r^3M(t)$
\cite{misner}.
In the case of dust ($p=0$) we have $M(t)=M_0$ and the amount of matter within the co-moving radius $r$ remains unchanged during collapse. In the case of perfect fluid we must set an initial condition for $M$ as $M(0)=M_0$. Then from the behaviour of $M(t)$ we see that during collapse there can be an inflow or an outflow of matter across the shell $r$.
Collapse is described by the adimensional scale factor $a(t)$, that is related to the physical area-radius $R$ by $R(r,t)=ra(t)$. Therefore once an initial scaling condition at $t=0$ is chosen (in our case we shall set $a(0)=1$) collapse proceeds as long as $\dot{a}<0$.
Then the metric is written as
\be
ds^2=-dt^2+\frac{a^2}{1-br^2}dr^2+r^2a^2d\Omega^2 \; ,
\ee
where $d\Omega^2$ is the line element on the unit two-sphere and $b$ is an integration constant that can be thought of as a condition imposed on the initial velocity of the particles. For the sake of clarity, in the following we will restrict our attention to the case of marginally bound collapse given by $b=0$.

Homogeneous models can be thought of as representing the inner core of the collapsing object, with inhomogeneities becoming more important as one moves away from the center. Therefore an increasing mass function $M$ implies that the outer shells are falling onto the inner shells.
Matching conditions should be imposed at the boundary of the cloud $r_b$, where the star's surface matches with a known exterior metric. In the following we shall assume that the homogeneous approximation is valid only in the vicinity of the core and that radial inhomogeneities will change the density and pressure profiles at greater radii.
Therefore in the following analysis we will not concern ourselves with the matching conditions at the boundary of the star. Note however that matching across a (possibly varying) boundary surface $r_b(t)$ with a generalized Vaidya exterior is always possible
(see \cite{matching}).

Also, in the present model we limit ourselves to isotropic pressures. The question whether anisotropies become important towards the formation of
the singularity was first addressed by Belinskii, Khalatnikov, and Lifschitz (BKL)
\cite{BKL}. 
During collapse, at least initially, spatial derivatives are less dominant with respect to time derivatives and anisotropies can be neglected. 
The BKL conjecture states that as one approaches the singularity a regime is reached where anisotropies dominate. However the BKL scenario is a
conjecture related to classical singularities in GR and it is not clear how it would translate in a quantum framework
(see for example
\cite{bojowald2}).
Diverging curvature is a fundamental part of the conjecture and being close to the singularity is the key ingredient so that spatial gradients can dominate. Therefore in bouncing scenarios, where the singularity is never reached, the role of anisotropies may become less important. 
In the following we will assume that as collapse proceeds we can always find a radius small enough so that anisotropies can be neglected.

The density and pressure of the cloud are given through Einstein's equations by
\be\label{rho}
\rho=\frac{3M}{a^3}, \; \; p=-\frac{\dot{M}}{a^2\dot{a}} \; ,
\ee
where dotted quantities denote derivatives with respect to $t$, and the remaining equations reduce the system to two differential equations for $M(t)$ and $a(t)$. Given the monotonic behaviour of $a$ it is always possible to use $a$ as a variable in place of $t$ and invert the equations to obtain $M(a)$ and $t(a)$. The differential equation for $M$ comes from the equation of state, by using equations \eqref{rho}, and takes the form
\be\label{M-cl-lin}
\frac{dM}{da}=-\frac{3kM}{a} \; .
\ee
Note that $M$ decreasing in $a$  corresponds to $M$ increasing in $t$. Then when $k>0$ the pressure is positive and diverges as collapse approaches the singularity.
The differential equation for $a$, that is the true equation of motion for the system, comes from the Misner-Sharp mass equation that in the marginally bound case reduces to $M(t)=a(t)\dot{a}(t)^2$ and can be written as
\be\label{a-cl-lin}
\frac{dt}{da}=-\sqrt{\frac{a}{M}} \; ,
\ee
with the minus sign chosen in order to describe collapse.
The classical collapse scenario has several features that from a physical point of view are not desirable. Most notably, the scale factor $a=(1-3(k+1)\sqrt{M_0}t/2)^{2/3(k+1)}$ reaches zero size in a finite time $t_s=2/3(k+1)\sqrt{M_0}$, thus indicating the occurrence of a space-time singularity where energy density and pressure diverge.

\section{Hagedorn phase}\label{hagedorn}
A first step to improve on the classical fluid model is to introduce an e.o.s. that saturates the number of states as the density increases.
In string theory and high energy physics the term Hagedorn temperature $T_H$ is used to indicate the temperature corresponding to a stage where ordinary matter is forced to convert into quark matter. At this point as energy is added to the system collisions between hadrons no longer increase the temperature but they create more and more particles. Then new quark-antiquark pairs can be spontaneously generated from vacuum thus providing arbitrary new degrees of freedom to the system. In a sense one can think of a system reaching the Hagedorn phase as allowed to store any arbitrary amount of energy without further increasing its temperature. In this sense the Hagedorn temperature associated with this state is the maximum temperature that can be reached in principle by matter
(see for example \cite{hagedorn}).
Measurements of neutron stars masses suggest that Hagedorn type equations of state are ruled out for neutron stars but they may still be valid for more dense exotic compact objects
(see for example \cite{ruffini}).
Also, the Hagedorn e.o.s. has been investigated in cosmological models
(see for example \cite{cosmology}) in connection with cyclic universe and the possible existence of primordial black holes.
Within classical collapse models there exist some approaches to relativistic collapse of Hagedorn fluids in the Vaidya space-time
(see for example \cite{harko}).
However classical collapse of a fluid sphere with an equation of state of this kind has not been investigated before.
Close to the Hagedorn phase the pressure increases less and less regardless of how much energy is added to the system. Then the fluid's heat capacity diverges at the temperature $T_H$ indicating that $T_H$ is a limiting temperature and it can be reached by the system only by providing an infinite amount of energy.

\begin{figure}[hhhh]
\includegraphics[scale=0.3]{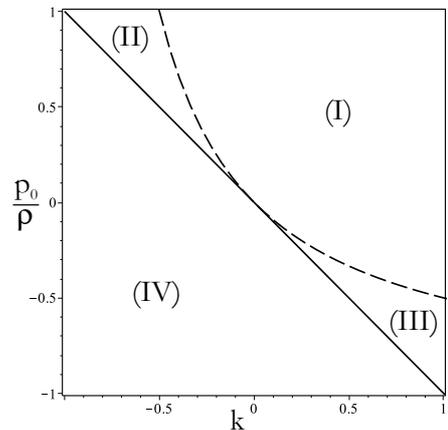}
\caption{Plot of $p_0/\rho$ as function of $k\in[-1,1]$. The weak energy condition $\rho+p\geq 0$ is satisfied in regions (I) and (IV) and violated in regions (II) and (III). The solid line is given by $-k$ while the dashed line is given by $-k/(k+1)$ and for $\rho=\rho_0$ represents the region where collapse halts and $a_{\rm cr}=0$.
}
\label{fig-wec}
\end{figure}

Such a situation may be effectively represented classically by the choice of an equation of state of the form
\be\label{hageos}
p=\frac{k\rho}{1+k\rho/p_0} \; .
\ee
The parameter $p_0$ describes the maximum pressure that can be achieved by the system and $p\rightarrow p_0$ as the density increases to infinity.
Typically for an astrophysical object one can think that the above equation of state will become important above nuclear densities, namely for $\rho$ greater than $10^{14}\text{gr}/\text{cm}^3$.
This equation of state has been used in
\cite{eos}
to describe a free bosonic string in Minkowski space in the context of quantum-gravity
but it has not been used in collapse scenarios.
Note that we recover the linear equation of state in the low density limit (if we take $p_0$ going to infinity).
Also it is worth noting that the equation of state is `soft' as the speed of sound $c_s$ within the cloud goes to zero as the density increases.
This is in contrast with other theoretical approaches to high density matter, such as the Zeldovich e.o.s., for which the fluid becomes stiff with $p\rightarrow\rho$ and the speed of sound tends to the speed of light as $\rho$ increases.
Weak energy conditions are always satisfied if we consider $k$ and $p_0$ positive. In the case where $p_0<0$ the e.o.s. is still physically reasonable if $k$ is negative. In this case w.e.c. are also satisfied for $k\geq -1$.
On the other hand if $k$ is negative (positive) and $p_0$ positive (negative) the e.o.s. does not have a clear physical interpretation since $p$ diverges as $\rho\rightarrow -p_0/k$. However one may be still tempted to consider the e.o.s. at large densities where $p$ tends to $p_0$ from above (below) as the density goes to infinity. In this case the w.e.c. are satisfied when $p_0\geq -k\rho/(k+1)$ and $p_0\geq-k\rho$ (respectively when $p_0\leq -k\rho/(k+1)$ and $p_0\leq-k\rho$, see figure \ref{fig-wec}).

By solving Einstein's equations for a fluid with the above e.o.s. we obtain a set of two differential equations.
The equation for the scale factor $a$ is again given by equation \eqref{a-cl-lin}, while the equation for the mass function $M$ becomes
\be\label{M-cl-Hag}
\frac{dM}{da}=-\frac{3kMa^2}{a^3+3kM/p_0} \; .
\ee
The above equation can be integrated to obtain $M(a)$ implicitly via
\be\label{M}
\left(\frac{M_0}{M}\right)^{1/k}=\frac{3kM+p_0(k+1)a^3}{3kM_0+p_0(k+1)} \; .
\ee
It is easy to see that the collapse scenario resulting from the choice of the equation of state \eqref{hageos} is very similar to the classical case with linear equation of state. In the case where $k$ and $p$ are positive it forms a singularity that is covered by an horizon at all times.
This is reasonable in light of the fact that no repulsive interactions are introduced as the density increases indefinitely.
The main difference with the linear e.o.s. model then is in the time of formation of the singularity that is now delayed (since pressures in this case are lower).
As said, the e.o.s. given in equation \eqref{hageos} is suitable also to describe negative pressures when $k<0$. In this case it is worth asking if it is possible to construct a model that collapses to a finite compact remnant. The condition for the cloud to settle to an equilibrium configuration is given by $\dot{a}=\ddot{a}=0$. In this case from
\be
\ddot{a}=-\frac{a}{2}\left(\frac{\rho}{3}+p\right) \; ,
\ee
we see that, as $\rho$ grows, $p$ must tend to $-\rho/3$ which gives the equilibrium condition as $(3k+1)p_0+k\rho_0=0$. Then for $k\leq-1/3$ it is possible to construct models for which $\ddot{a}$ goes to zero. However in order to have also $\dot{a}\rightarrow 0$ we must require $M\rightarrow 0$ and from equation \eqref{M} we see that this implies that $a\rightarrow 0$. Therefore no finite size compact remnant can be constructed in the classical Hagedorn collapse model.

\section{Semiclassical effects}\label{semi}
As it is well known, collapse does not halt in classical scenarios where energy conditions are always satisfied
and the attractive nature of gravity leads inevitably to the formation of a space-time singularity.
Nevertheless one can expect that quantum corrections, that should appear in the strong field, will cause
repulsive effects that may prevent the cloud from collapsing to a singularity.
Several approaches in this direction have shown that taking these effects into account leads to a bouncing scenario
where the cloud reaches a minimum size and then re-expands indefinitely
\cite{semiclass}.
The main idea is to treat the modifications to general relativity that must occur in the strong field limit as
an effective matter source to be added to the energy momentum tensor
\cite{barcelo}.
This way one solves the usual Einstein's equations for
an unphysical matter distribution (the effective energy-momentum tensor) that takes into account the modifications in the strong field.
The effective density $\rho_{\rm eff}$ and effective pressure $p_{\rm eff}$ can thus violate energy conditions as they are not the
physical density and pressure. One promising approach in this direction comes from Loop Quantum Gravity and involves quadratic corrections to the energy density as $\rho$ approaches a critical value $\rho_0$
\cite{LQG}.
In the semiclassical formalism this implies taking an effective density $\rho_{\rm eff}$ given by
\be
\rho_{\rm eff}=\rho\left(1-\frac{\rho}{\rho_0}\right) \; ,
\ee
where the parameter $\rho_0$ describes the maximum density that can be achieved by the system and signals the regime where quantum effects cannot be neglected.
Typically for a star one can think that quantum effects will become important around the Planck scale (namely for $\rho>10^{94}\text{gr}/\text{cm}^3$).
Therefore by taking $\rho_0$ of the order of the Planck density we are constructing a semi-classical description of quantum-gravitational effects coming from a first order approximation of collapse in LQG.
Note that the energy scale of quantum corrections is several orders of magnitude higher than that of the Hagedorn phase.
However the constant $\rho_0$ is model dependent, as it comes from the specific approach chosen to deal with repulsive effects, and need not necessarily be of the order of the Planck density. Then the above formalism may still be used with different values for the density parameter $\rho_0$ coming from a different theoretical approaches.
For example, in
\cite{Antonino}
it was suggested that four-fermion interaction may halt collapse before the quantum-gravity regime and thus allow for the existence of compact objects.
This approach has been used in
\cite{BMM}
to describe semiclassical dust collapse.
The choice of the effective density in turn leads to an effective pressure and an effective mass function as given by
\bea
M_{\rm eff}&=&M\left(1-\frac{\rho}{\rho_0}\right) \; , \\
p_{\rm eff}&=&p\left(1-\frac{2\rho}{\rho_0}\right)-\frac{\rho^2}{\rho_0} \; .
\eea
The differential equation that must be satisfied by the mass function $M$ is again given by equation \eqref{M-cl-lin}, while the equation for the scale factor $a$ now becomes $\dot{M}_{\rm eff}=a\dot{a}^2$ that can be written as
\be\label{a-semi-lin}
\frac{dt}{da}=-\sqrt{\frac{a}{M}\frac{1}{\left(1-\frac{3M}{\rho_0 a^3}\right)}} \; .
\ee
Note that as $\rho_0$ goes to infinity we recover the classical scenario. As mentioned, the final results is that the cloud reaches a minimum scale $a_{\rm cr}$ and then bounces back. At the time of the bounce the effective density and the effective mass function become zero and thus quantum effects counterbalance the classical gravitational attraction effectively `turning gravity off'. This limit can be viewed as the semiclassical equivalent to approaching asymptotic freedom
\cite{asymtotic-s}.

We now turn the attention to the semiclassical scenario for a fluid that has reached the Hagedorn phase.
This means considering the mass function given by \eqref{M-cl-Hag} and the scale factor from \eqref{a-semi-lin}.
The existence of a threshold near the Planck density for which modifications to
the classical collapse scenario are necessary is not a new idea
(see for example \cite{2scales}).
However, note that now there are two different scales at which corrections to the classical perfect fluid model become important, namely the Hagedorn phase, determined by $p_0$, and the quantum-gravity phase, determined by $\rho_0$.

\begin{figure}[hhhh]
\includegraphics[scale=0.3]{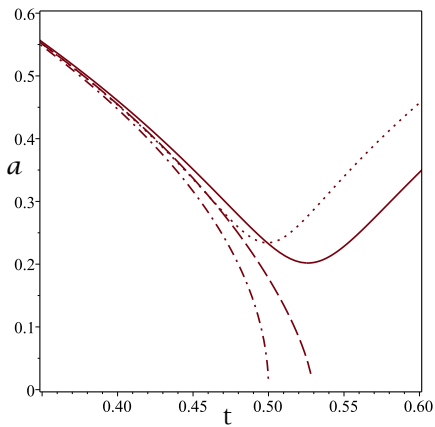}
\caption{Comparison of the scale factor in the four cases of classical or semiclassical collapse with linear e.o.s or Hagedorn e.o.s. with the parameters chosen as $k=1/3$, $p_0=50$ and $\rho_0=1000/3$ and the initial conditions taken as $a(0)=1$, $M_0=1$. (i) Solid line is semiclassical Hagedorn collapse, (ii) dotted line is semiclassical collapse with linear e.o.s., (iii) dashed line is classical Hagedorn collapse and (iv) dotted-dashed line is classical collapse with linear e.o.s.. Note that cases (i) and (ii) lead to a bounce when $a=a_{\rm cr}$, while cases (iii) and (iv) lead to the formation of a singularity when $a=0$.}
\label{fig1}
\end{figure}

It is not difficult to see that when $k>0$ and $p_0>0$ the semiclassical model results in a bouncing scenario. In fact we see that as $a\rightarrow a_{\rm cr}$ the collapsing velocity $\dot{a}$ goes to zero but the acceleration $\ddot{a}$ does not vanish, thus indicating the occurrence of the bounce.
However it should be noted that the possibility that the bounce leads to the formation of a baby universe may not
be excluded in principle. In this case the outer event horizon would remain unchanged, thus giving rise to a black hole
for observers at infinity, while the expanding matter would be confined within the newly formed black hole.
Such a scenario would require a phase transition for the
collapsing matter to occur at the moment of the bounce. Mathematically this translates
into the requirement that some suitable matching conditions be satisfied at the surface
given by $t=t_B$.
Such analysis is beyond the scope of this article and it will
be carried out elsewhere. On the other hand in the bouncing scenario the black hole turns into a
white hole after the bounce and the expanding solution is simply described by the time reversal of the collapsing one.
In this case it is worth asking under which conditions a compact remnant may form.
After a straightforward calculation we see that as $a$ approaches the critical value $\ddot{a}$ tends to the value
\be
\ddot{a}_{\rm cr}=\frac{a_{\rm cr}}{2}(\rho+p).
 \ee
 Therefore the fluid model must approach the behaviour of a `dark energy' fluid with $p=-\rho$ for the repulsive effects to halt collapse. Note that the condition for equilibrium is more stringent in the semiclassical model with respect to the classical case. For the e.o.s. considered here we get
\be
\ddot{a}\longrightarrow\frac{a_{\rm cr}\rho_0}{2}\left(1+\frac{kp_0}{p_0+k\rho_0}\right) \; ,
\ee
from which we see that the condition $\ddot{a}=0$ can not be satisfied also in the case where $-1<k<0$ when $p_0<0$.
We conclude that obtaining $\ddot{a}=0$ is possible only by imposing $k<0$ and $p_0>0$ or $k>0$ and $p_0<0$.
However, the behaviour of the equation of state in these two cases is not physically very meaningful.
Nevertheless, for the sake of argument, let us now focus for a moment on the case $k<0$ with $p_0>0$.
In this case the e.o.s. \eqref{hageos} can be considered valid only at high densities and $p$ approaches the limiting value $p_0$ from above as $\rho$ goes to infinity.
If we want to satisfy the equilibrium condition we must choose the value of $p_0$ to be $p_0=-k\rho_0/(1+k)$, so that, as $\rho$ grows approaching the critical value, the pressure balances the attraction giving $\ddot{a}=0$ (see figure \ref{fig-wec}).
Note that the above constraint, in order to have the value of $p_0$ several orders of magnitude lower than that of $\rho_0$, implies that $k$ must be small.
Conversely, assuming to know the values of $p_0$ and $\rho_0$ we can evaluate the value of $k$ for which collapse halts.
With the above choice of $p_0$ we can achieve the equilibrium configuration, but from equation \eqref{M} we see that this implies that $a_{\rm cr}\rightarrow 0$.
Therefore we conclude that no compact remnant can be constructed in the semiclassical Hagedorn collapse model.

However other repulsive effects, not due to gravity, may contribute to create an exotic object leftover from collapse, much in the same way as neutron degeneracy pressure balances gravity in neutron stars.
Then from the fact that the scale factor reaches the limiting value given by
\be
a_{\rm cr}^3=\left(\frac{3M_0}{\rho_0}\right)^{\frac{1}{k+1}}\left(\frac{3kM_0+(k+1)p_0}{k\rho_0+(k+1)p_0}\right)^{\frac{k}{k+1}} \; ,
\ee
we see that the critical density $\rho_0$ plays a crucial role in determining the size of the final remnant.
The bigger the value of $\rho_0$ the smaller the critical scale of collapse.

\begin{figure}[ttt]
\includegraphics[scale=0.3]{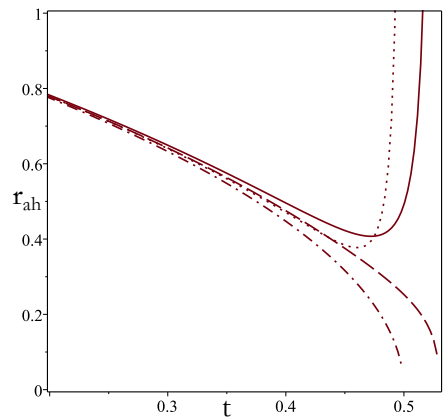}
\caption{Comparison of the apparent horizon in the four cases of classical or semiclassical collapse with linear e.o.s or Hagedorn e.o.s. with the parameters chosen as $k=1/3$, $p_0=50$ and $\rho_0=1000/3$ and the initial conditions taken as $a(0)=1$, $M_0=1$. (i) Solid line is semiclassical Hagedorn collapse, (ii) dotted line is semiclassical collapse with linear e.o.s., (iii) dashed line is classical Hagedorn collapse and (iv) dotted-dashed line is classical collapse with linear e.o.s.. Note that in cases (i) and (ii) the apparent horizon curve $r_{\rm ah}$ goes to infinity as $a\rightarrow a_{\rm cr}$, while in cases (iii) and (iv) $r_{\rm ah}\rightarrow 0$ as $a\rightarrow 0$.}
\label{fig2}
\end{figure}

\begin{widetext}

\begin{figure}[h]
\centering
\begin{minipage}{.45\textwidth}
\centering
\includegraphics[scale=0.3]{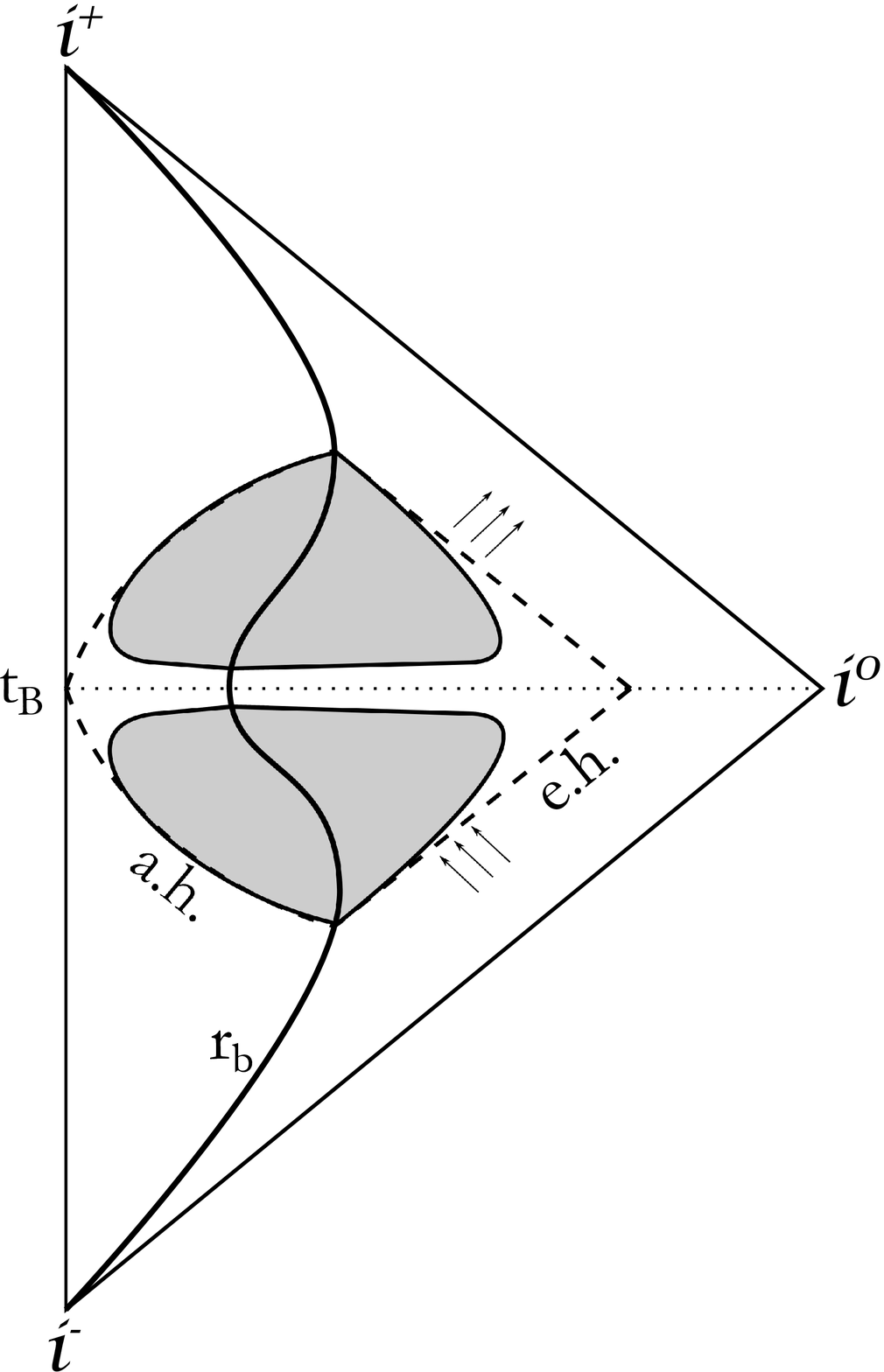}
\end{minipage}
\hfill
\begin{minipage}{.45\textwidth}
\centering
\includegraphics[scale=0.3]{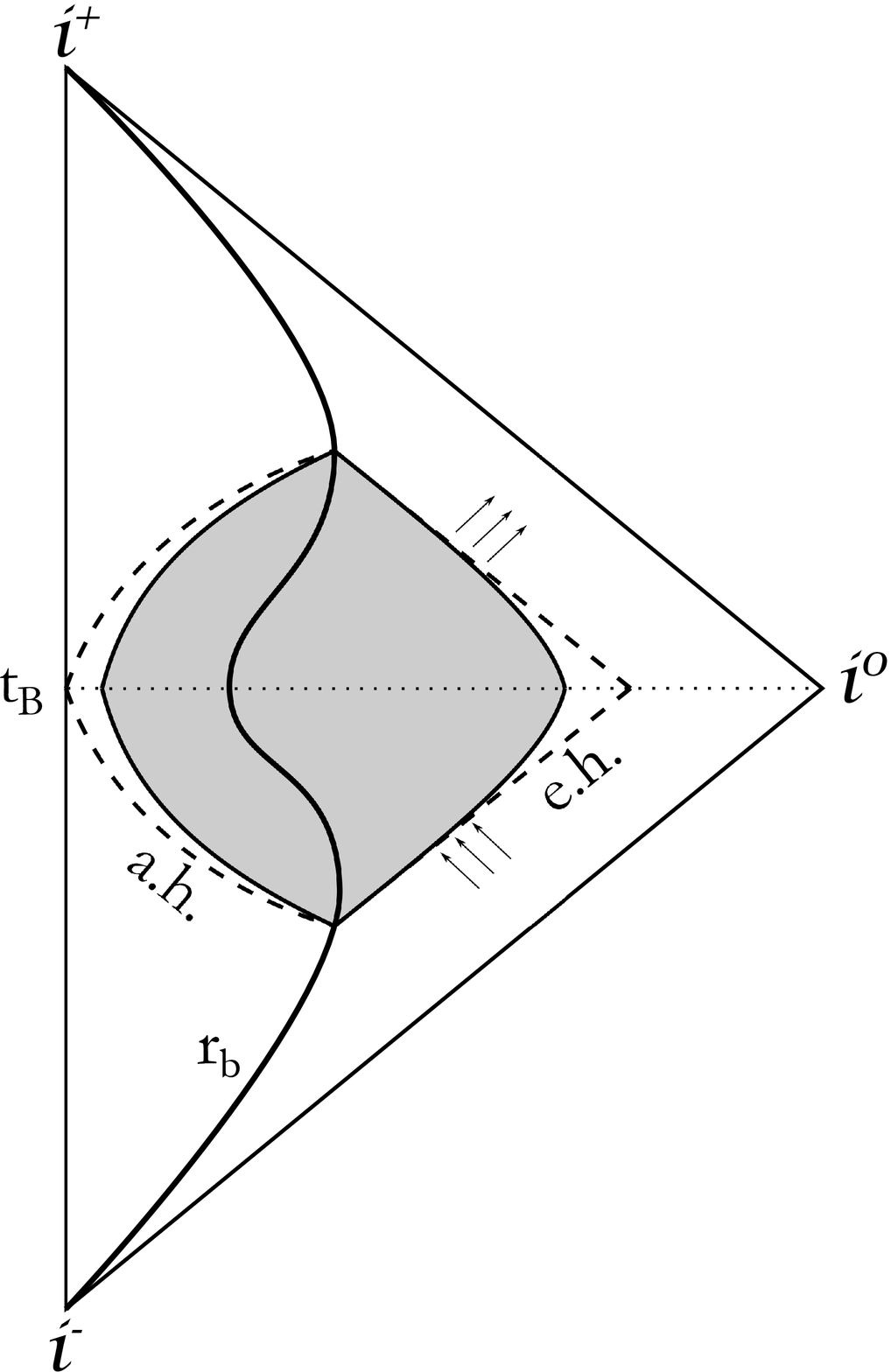}
\end{minipage}
\caption{The thick line represents the boundary of the cloud $r_b$, the grey area represents the trapped region. Dashed lines represent the classical apparent horizon (a.h.) and event horizon (e.h.), without semiclassical corrections. After the time of the bounce $t_B$ the cloud expansion is described by the time reversal of the collapsing solution. Left Panel: Penrose diagram of homogeneous collapse with semiclassical corrections. Every shell bounces at the same time $t_B$ when the effective mass becomes zero. Right panel: Penrose diagram for the more realistic case of inhomogeneous collapse. As the central shell reaches the critical density the outer shells still have non zero effective mass. The total mass of the system decreases until the time of the bounce and then increases again. The outer horizon shrinks to a minimum radius due to the decrease of the effective mass. After the bounce the black hole turns into a white hole.}
\label{fig-pen}
\end{figure}

\end{widetext}

The main consequence for astrophysical black holes can be inferred from the study of the behaviour of the apparent horizon in the interior of the collapsing cloud.
The condition for the formation of trapped surfaces in the classical scenario is given by
\be\label{ah}
1-r^2\frac{M}{a}=0 \; ,
\ee
that gives the time at which the shell $r$ becomes trapped. This can be expressed via the apparent horizon curve $r_{\rm ah}(t)=\pm\sqrt{a(t)/M(t)}$. We see that, as $a$ goes to zero, if $M$ goes to a constant (or goes to zero slower than $a$) then $r_{\rm ah}$ goes to zero as well.
The same can be seen from $M=a\dot{a}^2$, since we can write $r_{\rm ah}(t)=1/\dot{a}$ that goes to zero as $\dot{a}$ diverges.
On the other hand in the semiclassical models the condition for the formation of trapped surfaces is
\be\label{ah-sc}
1-r^2\frac{M_{\rm eff}}{a}=0 \; ,
\ee
where now $a$ reaches a minimum value at which $\dot{a}=0$ and $M_{\rm eff}=0$.
Then it is easy to see that in this case the radius of the apparent horizon must go to infinity as the scale factor reaches the critical value $a_{\rm cr}$.
This means that $r_{\rm ah}$ will cross the boundary of the star at some point before the critical scale is reached, thus leaving the physics that occur for $a$ close to $a_{\rm cr}$ not covered by any horizon (see figure \ref{fig2}).
However this feature may be due to the fact that we are neglecting inhomogeneities. In the more realistic case of inhomogeneous collapse the outer shells still gravitate as the central shell reaches the critical density. As a consequence the horizon does not disappear and the black hole turns into a white hole after the bounce (see for example
\cite{LMMB}). 
In figure \ref{fig-pen} is shown the comparison between Penrose diagrams for collapse in these two cases.
Note that if the boundary of the collapsing core is smaller than the minimum value of the apparent horizon curve, $a_{\rm cr}<{\rm min}(r_{\rm ah})$, then the collapsing object may not be covered by the horizon at any time (depending on what happens to trapped surfaces in the outer layers).

\section{Discussion} \label{disc}
We have constructed a simple analytical toy model for relativistic collapse that includes a reasonable equation of state for high density matter as well as modifications to classical general relativity in the strong field regime. In this model the pressure of the system has an upper limit related to the Hagedorn temperature and the asymptotic safety regime is achieved at high densities.
This implies that there are two scales as determined by the values of two parameters (namely the maximum pressure and the maximum density).
We have shown that under these circumstances no singularity is produced and collapse reaches a minimum size after which the cloud re-expands.
Models such as the one presented here suggest that black holes, as defined mathematically in terms of singularity covered by an event horizon at all times, may not exist in nature. The only horizons that may occur in realistic scenarios are apparent horizons and these must be viewed as transient phenomena
\cite{hawking}.
In the present model the black hole formation scenario turns into a white hole after the bounce.
The most important consequence for astrophysics is that the physics of collapsing stars may be hiding quantum-gravitational effects at its core.
And these effects may have detectable observational signatures.
It has been suggested that such bouncing models may appear as powerful explosions in the universe
\cite{rovelli}.
Therefore, if such hypothesis were to be confirmed, these models would provide a cosmic laboratory to study quantum-gravity, thus allowing us
to probe energy scales that cannot be reached on Earth.

Further to this, the existence of a minimum size at which collapse stops suggests the possibility of the existence of exotic compact remnants.
Such objects would have to be smaller, denser and less massive than a neutron star and they would be intrinsically quantum in nature.
Note that the final size of such an object is related to the value of the maximum density parameter which may or may not be related to the Planck scale regime.
At this point one is naturally led to wonder how such an object can be detected, if it exists, and what kind of observational features it would have.
Studies of accretion disks around classical solutions with naked singularities indicate that the luminosity emitted by a disk around an exotic compact object may be higher than that emitted by the accretion disk around a black hole
\cite{obs}.
Nevertheless this feature may be due to the presence of a singularity in the classical solution while the true nature of the real objects may be entirely different. Detection of observable phenomena coming from exotic compact objects will be a challenge for future experiments, both in terms of the strength of the signals as well as the statistics of their occurrence. Nevertheless, future observations of strong gravity phenomena, via multi-messenger astronomy, will hopefully provide the much needed experimental data to test general relativity in the strong field and to put to test the various hypothesis regarding the final fate of collapse of very massive stars.


\end{document}